\documentstyle[12pt]{article}

\newcommand{\be}{\begin{equation}}
\newcommand{\ee}{\end{equation}}
\newcommand{\ba}{\begin{eqnarray}}
\newcommand{\ea}{\end{eqnarray}}
\newcommand{\no}{\nonumber \\}
\newcommand{\half}{\frac{1}{2}}
\newcommand{\abs}[1]{\mid{#1}\mid}
\begin{document}
\begin{titlepage}
\pagestyle{empty}
\vspace{1.0in}
\begin{flushright}
hep-th/yymmxxx\\
\end{flushright}
\vspace{.1in}
\begin{center}
\begin{Large}
{\bf The Super-potential and Holomorphic properties of the MQCD}
\end{Large}
\vskip 0.5in
Sang-Jin Sin
\vskip 0.2in  
{Department of Physics, Hanyang University,
Seoul 133-791, Korea\\
{\tt sjs@dirac.hanyang.ac.kr}
}
\end{center}
\vspace{0.5 cm}

\begin{abstract}
We study the holomorphic properties of the MQCD by comparing 
the super-potentials  in MQCD and the gauge theory.
First we show that the super-potential defined as an integral of three form
is NOT appropriate  for generic situation with  quarks.   We report a resolution  
of the problem  which works for the brane configurations of 90 degree
rotation, including the true SQCD. The new definition  does not need 
auxiliary  surface and can be reduced to a contour integral for some cases. 
We  find relation beetween the new and old definitions, which is 
verified by explicit calculation  for SU(N), SO(N), Sp(N)  simple groups
with $F$ of massive  quarks.
\end{abstract}

\end{titlepage}

\section{Introduction}
The idea of D-brane\cite{pol} opened up a new and surprisingly simple way to
communicate between super symmetric (SUSY) gauge theories and the
 string theories\cite{hw,egk,egkrs,witten1,oz1,yanki1,witten2}. In a recent paper
\cite{witten1}, Witten provided solutions of $N=2$ SUSY gauge theories in
four dimension\cite{sw} by reinterpreting web of branes in type II 
superstring as a single  M-brane. In a subsequent paper\cite{witten2}, 
he showed how some of the long standing
problems in particle physics such as quark confinement and chiral symmetry
breaking, can be approached from the M-theory point of view. There, he also
suggested a way to calculate the super-potential (SP) for MQCD, which 
immediately gives the tension of the
domain wall \cite{cs}.  Although these are examples of spectacular successes,
the reason why it work is not very clear. In fact the region where M-theory 
is working is quite opposite limit to the region where the gravity is 
decoupled\cite{ooguri}. In fact, 
there are evidences \cite{witten2,oz3} that MQCD and the SQCD  is not the 
same theories.  However, 
it is speculated \cite{oz2,oz3} that the holomorphic objects of two theories may be 
the same. So it is compelling to check whether the holomorphic properties of 
two theories really agree.

In this paper we study these issues by studying the super-potential for various 
M-brane configurations and  comparing with corresponding gauge theories.
In the  previous work  of the author \cite{sin},  the Witten's calculation of the super 
potential was  generalized to the case where gauge group is the  product of 
$SU(k_i)$`s as well as to the case where massless quarks are included.  
There, it was claimed that the value of  $W_3$, the super-potential in MQCD 
defined in \cite{witten2}, is the same as that of the gauge theory with  massive 
adjoint fields. 
It turns out that this agreement is limited  to   very specific cases and also 
very limited sense that will be described in detail later. 
Main observation of this paper is that $W_3$ should be replaced by a new definition 
in order to compare MQCD with SQCD in the presence of matter.

In section 2, we give a review to set up the language.
In section 3, we  first consider the cases for $SU(N)$ gauge theory with 
massive adjoint and  hyper multiplets and 
clarify the staement made in the previous work \cite{sin}.
Namely, in what sense, $W_3$  agrees with gauge theory for 
non-parallel but non-orthogonal NS five branes. 
Then we point out that  $W_3$ for the 
orthogonal NS five branes, which corresponds to  the true  N=1 SQCD with 
hyper multiplets,  disagrees with the minimum value of the gauge theory super-potential.
In section 4,  we give a modified  definition of SP, $W_2$, which  
does agree with SQCD. The new definition is also ambiguity free since 
it does not need auxiliary  surface.
We will  find relation beetween the new and old definitions through these 
calculations.  
In section 5, we  will give   more examples  for  the theory  with SU(N),
SO(N), Sp(N)  gauge groups with $F$ of massive  quarks.

\section{Review of the Super-potential in MQCD}
Let $\Sigma$ be a surface that describes the 
five brane  as ${\bf R}^4 \times \Sigma$. The $\Sigma$ is 
embedded in the six dimensional internal space-time, 
$x^4,x^5,x^6,x^{10},x^8,x^9$. Let $v,w,t$ be  the complex 
co-ordinates defined as $v=x^4+ix^5, \; w=x^8+x^9,\; 
t=e^{-(x^6+ix^{10})/R}$, where $R$ is the size of eleventh dimension.
From the earlier study\cite{yanki1,oz1}, we know that $v\sim <\Phi>$ hence 
we assume that $v$ has mass dimesion one and 
$ w\sim <\tilde{Q}Q> $ hence mass dimension two.
Consider a holomorphic top form in the complex three dimensional space 
whose co-ordinate are $v,w,t$:
\be
\Omega_3= dv\wedge dw \wedge dt/t .
\ee
Given a brane configuration, the super-potential must give a definite value. 
The only  canonical  candidate is integral of the $\Omega_3$ 
over a three manifolds. To provide such a manifold, Witten introduced 
an auxiliary surface $\Sigma_0$ in the same homology class
of $\Sigma$ such that $\Sigma_0$ is asymptotic to the $\Sigma$.  
Then there exists a three manifold $B$ enclosed by $\Sigma$ and $\Sigma_0$
and  the super-potential is defined by 
\be 
W_3(\Sigma,\Sigma_0)= W(\Sigma) - W(\Sigma_0) = {1\over \pi i} 
\int_B \Omega_3.\label{w3def}
\ee  
This defines $W(\Sigma)$ up to an additive constant.
\footnote{It is very similar to the Wess-Zumino term in field theory in the sense 
that it require one more dimension. }
We included the ${1\over \pi i}$ factor since it is the right normalization 
when we compare with the gauge theory results.
Notice that we do not have the factor $R$ in the above definition unlike 
ref.\cite{witten2}, due to our mass dimension of $v,w$. 
Namely, the super-potential must have 
mass dimension three and the sum of the mass dimension of $v$ and $w$ is already three,
hence there is no room for $R$. The super symmetry R charge is carried by $v$. 
 In order to assign a unique value $W(\Sigma)$ 
to a given brane configuration 
$\Sigma$,  $W(\Sigma_0) $ must be universal. In order to achieve this goal,
Witten required $\Sigma_0$ be invariant under chiral rotation
so that one can put $W(\Sigma_0) =0$.  In case this is impossible,  
we  have to regards that $W_3$ depends on the pair of surfaces $\Sigma,\Sigma_0$.

\section{ Difficulties of $W_3$}
First  we consider MQCD corresponding to $SU(N)$ gauge theory  with $F$ (massive) quarks 
as well as quadratic super potential  for the adjoint field $\Phi$.
The curve $\Sigma$ is given \cite{oz1} in parametric form by . 
\begin{eqnarray}
v(\lambda) & = &-m_f +{1\over\mu}{(\lambda-\lambda_+)(\lambda-\lambda_-)\over\lambda}                                    \\
w(\lambda) & = & \lambda           \label{finitemu}\\
t(\lambda) & = & \lambda^{N-F} (\lambda-\lambda_+)^r(\lambda-\lambda_-)^{F-r}
. \no
\end{eqnarray}
To construct a three volume, we need an auxiliary curve $\Sigma_0$ which we
define by 
\begin{eqnarray}
v & = &\left( {\lambda_+\lambda_-\over\mu\lambda} -m_f \right)
f_0(\abs{\lambda})
  +\left({\lambda-\lambda_+-\lambda_- \over\mu}\right)  f_\infty(|\lambda|) \no
w  & = & \lambda f_\infty (|\lambda|)  \label{sigma0} \\
t  & = & \lambda^{N-F} (\lambda-\lambda_+)^r(\lambda-\lambda_-)^{F-r}
,   \no
\end{eqnarray}
where $f_{0}(|\lambda|)=1$ for $|\lambda |<\epsilon$ for small enough $\epsilon$,
and $f_{\infty}(|\lambda|)=1$ for $|\lambda |> \half {\rm min}(|\lambda_-|,|\lambda_+| )$ 
and vanish rapidly outside.  The subtlety and problems in this choice will be 
discussed after the calculation. 
Now $ B$ is defined as the volume interpolating $\Sigma$ and $\Sigma_0$.
It can be parameterized by $g_\alpha(|\lambda|,\sigma)$ 's 
which interpolate  $f_{\alpha}$ and 1 as $\sigma$ varies from 0 to 1.
\begin{eqnarray}
v & = &\left( {\lambda_+\lambda_-\over\mu}{1\over \lambda} -m_f \right)
g_0(\abs{\lambda},\sigma)  +\left({\lambda-\lambda_+-\lambda_- \over\mu}\right) 
g_\infty(|\lambda|,\sigma)\\
w  & = & \lambda g_\infty (|\lambda|,\sigma)   \label{B}  \\
t & = & \lambda^{N-F} (\lambda-\lambda_+)^r(\lambda-\lambda_-)^{F-r}
,       \label{}
\end{eqnarray}
From this,  
\be
W_3:=W_3(\Sigma)-W_3(\Sigma_0) =-{\lambda_+\lambda_- \over \mu}(N-{1\over2}F) + 
{1\over2}m_f (r\lambda_+ +(F-r)\lambda_-) . \label{answer}
\ee
This is equal to the minimum value of 
\be
W_{{\rm {eff}}}= (N - F) \left(\frac{\Lambda _{N=1}^{3N - F}}{\det M 
}\right)^{1/(N -F)} +\frac{1}{2\mu } \left( {\rm Tr}(M^2)-\frac{1}{N }(
{\rm Tr}M)^2\right)  + {\rm Tr}(m_f M) \label{gaugepot},
\ee
if we identify the $\lambda_\pm$ with two distinguished eigenvalues of the 
meson matrix $M$.  
A technical remark: The appearance of weighting factor $\half$ in eq. \ref{answer} is 
a chracteristic feature of $W_3$. This factor gives the possibility to find a choice of 
$\Sigma_0$ that gives the gauge theory value when there is a quadratic super potential
${\rm Tr}\Phi^2$. 
In the limit $\mu \to \infty$ \cite{oz1}, $\lambda_+/\mu  \to - m_f$ and 
\be 
W\to N\zeta , \label{watortho}
\ee
where $\zeta=m_f\lambda_-$. This is equal to the 
minimum values of the gauge theory super-potential without $1/\mu$ term.

Apparently, this looks good. However, we should notice that the value of 
$W_3$ depends on  $\Sigma_0$  very sensitively.
$\Sigma_0$  is  not chiraly invariant and  in fact
  it is impossible to choose chiraly invariant  $\Sigma_0$ that is 
asymptotic to $\Sigma_0$ since the chiral symmetry is 
explicitly broken by the mass.  Therefore one can not set its super-potential zero and 
the value $W_3(\Sigma,\Sigma_0)$ can not be attributed to the brane
configuration only. This is the first difficulty.  
If we ask whether one can find a class of auxiliary surface $\Sigma_0$, 
then  from the above calculation the answer is yes. In fact the conclusion 
of ref.\cite{sin} should be considerred only in this sense.

In fact a much more natural choice of  $\Sigma_0$  defined as 
\begin{eqnarray}
v & = & v(\lambda) f_0(\abs{\lambda})\no
w  & = & w(\lambda) f_\infty (|\lambda|)  \label{sigma1} \\
t  & = & t(\lambda)   \no
\end{eqnarray}
fails to fit the gauge theory value.  Apart from  the simplicity and the generality
it satisfies all the asymptotic conditions 
by choosing $f_\infty(\abs{\lambda}) =1$ for 
$|\lambda| > {\rm max}(|\lambda_-|,|\lambda_+|)$ .
However,  according to the definition of $W_3$ it gives  
\ba 
W_3(\Sigma,\Sigma_0) ={\lambda_+\lambda_- \over \mu}(N- F) +{1\over2}m_f (r\lambda_+ +
(F-r)\lambda_-)  
\ea
which is different from the gauge theory value.

Now  we calculate  $W_3$ for  M-brane 
configuration that correspond to the 90 degree rotation ($\mu\to \infty$).
Namely,  if we take the limit $\mu \to \infty$ of the eq. \ref{finitemu}, we 
get \cite{oz1}
\begin{eqnarray}
        v & = & -\zeta/\lambda \no
        w & = & \lambda \no
        t & = & \lambda^{N-F} (\lambda-\lambda_-)^{F}
        \label{ortho},
\end{eqnarray}
Using the method described above, we get 
\be
W_3= \zeta(2N-F)
\ee
This  value is not consistent with the gauge theory value of eq.\ref{watortho}.
The same result is obtained for the configuration where not all the 
eigenvalues of the meson matrix are degenerate.
Therefore there is a discontinuity at $1/\mu=0$:
\be
\lim_{\mu \to \infty} W_3(\Sigma(\mu)) \neq W_3\left(\lim_{\mu \to \infty} 
\Sigma(\mu)\right) .
\ee
For  orthogonal 
NS branes $vw=\zeta$, so there is no room to play with in choosing the $\Sigma_0$,
therefore there is no $ \Sigma_0$  such that it is asymptotic to 
$\Sigma$  and  $ W_3 $ fits the gauge theory value. 
So it is much worse than the case of non-orthogonal NS branes.

Summarizing, 
Witten's  definition of the super-potential\cite{witten2}
involves an auxiliary surface 
and this cause a lot of difficulties for the case including  
(massive) quarks especially for orthogonal five branes.
The ambiguity in the super-potential coming from the dependence on 
$\Sigma_0$ was stressed both in \cite{sin} and \cite{oz2}.
Therefore it is reasonable to conclude that 
$W_3$  is not appropriate to use to compare with gauge theory for 
generic situation. So we need some other method to calculate the 
M-theoretic super-potential.
In the next section, we report a resolution  to the problem
which works for the brane configuration of 90 degree 
rotation, including the true SQCD.

\section{New definition of super-potential in MQCD}

We try to modify the  definition of the super 
potential such that it depends only on   $\Sigma$.
 Since $\Sigma$ is a surface, we look for
a 2-form $\Omega_2$ such that $d\Omega_2=\Omega_3$ to define 
\be
W_2 \sim \int_\Sigma \Omega_2.
\ee
For a closed   surface $\Sigma$, it  is just rewriting  by Stokes' theorem. 
However, since  $\Sigma$ is not a compact  manifold in our application,
there is an ambiguity corresponding to the gauge transformation:
\be \Omega_2\to \Omega_2+df_1.
\ee 
That is, for any one form  $f_1$,  $d(\Omega_2+df)=d\Omega_3 $.
But the integral of $df_1$ over the $\Sigma$ is non-zero 
since it is not a compact manifold.  This 
apparent ambiguity is resolved by noticing the symmetry of the 
$\Omega_{3}$.  It is antisymmetric under the exchange of 
$v$ and $w$.  Requiring the same  symmetry to 
$\Omega_{2}$ fixes the two form  $\Omega_{2}$ uniquely, 
\be 
\Omega_2=\frac{1}{2}(vdw-wdv)\wedge dt/t \label{omega2}.
\ee
Therefore it is reasonable to  take
\be
W_2={1\over  \pi i} \int_\Sigma \frac{1}{2} (vdw-wdv)\wedge dt/t \label{w2}
\ee
as the new definition of the super-potential:  it 
is independent  on the auxiliary surface $\Sigma_{0}$ and it respects many 
of the symmetry of $W_3$.  However,
the pull back of the holomorphic two form on a holomorphic
curve is simply zero.  Therefore
one should ask how to get non-zero value of the the super-potential.
 What saves us from the triviality
is precisely the non-compactness (singularity) of the Rieman surface. 
In the presence of the singularities, one should take
out small disks around them. Once the singularities are cut out,
the integrand becomes either identically zero or total derivative term, 
which in turn becomes a line integral over the sum 
of the  boundaries of the small disks.  In this way our definition reduces to
 a contour integral. 
  
We now illustrate these idea for the cases where NS and NS' branes are orthogonal.
First, take a pull-back of 
the  surface to the  $v$- or $w$- plane using the equations that define $\Sigma$:
\begin{eqnarray}
 w&=&w(v), \;\;t=t(v)  \; \hbox{for  the pull-back to v-plane } \no
  v&=&v(w), \;\;t=t(w) \; \hbox{for  the pull-back to w-plane } .
\end{eqnarray}
According to which pull-back we make, the algebraic form of the curve 
look different. 
It has been known \cite{oz1,yanki1} that  the co-ordinate $v$ corresponds to  
the adjoint  field $\Phi$ in the effective action  and $v$'s values
are  eigenvalues of the quark mass matrix $m_{ij}$ due to the coupling
${\tilde Q}\Phi Q$.
Similarly the co-ordinate $w$ corresponds to the meson fields
$M_{ij} =  {\tilde Q}_i Q_j$ and  $w$'s values  are eigenvalues of the 
meson matrix which again corresponds to the mass of the dual quarks via 
$\tilde{q}Mq$.  Therefore  it is appropriate to call the pull-back to the $w$-plane 
 $w$-  or meson-picture and call the pull-back to the v-plane  $v$- or 
 $\Phi$-picture \footnote{These are not yet the same as the 
`electric' and `magnetic' theory, since even for the  $N>F$ case, this
classification is valid.} 

In what picture should we work?
In a gauge theory, the super-potential was always evaluated in 
terms of the meson fields,  
hence one should work in meson-picture to compare the value of the super 
potentials of MQCD with that of SQCD.
In  meson picture,  $v(w)dw\wedge dt(w)/t$ is zero due to 
the holomorphic dependence  of $dt(w)$ on $w$. 
Hence  the super-potential becomes
\ba
W_2(\Sigma)&=& -{1\over 2\pi i} \int_{\mathbf R^{2}_{w}} wdv(w) \wedge dt(w)/t\no
                &=& - {1\over 2\pi i} \int_{\cup_j C_j} wv(w)dt(w)/t \no
                &=& {1\over 2\pi i}  \zeta \int_{C_\infty} dt(w)/t  \no
                &=&  \zeta( \hbox{No. of zeroes  - No. of poles of t})  
\ea 
where $C_j$'s are the small curves around the singularities of the curve $\Sigma$
and $C_{\infty}$ is the circle around the $w=\infty$. We have used  $vw=-\zeta$.
Therefore  the value of the super-potential, when NS and NS' brane are 
orthogonal,  is proportional to the asymptotic  bending 
power of NS' brane, defined  as $t\sim w^{Power}$ as $w\to \infty$. 

One should also notice that we cut out the disks around 
all the singularities of  the curves, namely those of $dt(w)/t$
as well as those of  $v(w)$.  
Therefore the tubes corresponding to the semi-infinite 
four branes  as well as   asymptotic circle of the NS' brane 
contributes to the super-potential directly.  As we will see in the 
explicit example below, the tubes' contribution is equal to the mass term 
${\rm Tr}{mM}$ in the gauge theory.

So far we suceeded to find an expression that is non-trivial and we want to see
that the precise relation between the super-potential in SQCD and MQCD is:
\be 
W_{SQCD}= W_2(\Sigma)
\ee
Later, we will verify  this relation for several examples which cover all the physically important
cases. For a moment, we will ask more general questions: 
what happen  if we define the super-potential in the v-picture as 
\ba
W^{v}_2(\Sigma)&=& \frac{1}{2\pi i}\int_ {\mathbf R^{2}_{v}} vdw(v) \wedge dt(v)/t \no 
                &=& \frac{1}{2 \pi i}\int_{\cup C_i} vw(v)dt(v)/t  \no
                &=&  -\frac{1}{2 \pi i}\zeta \int_{C_{v,\infty}} dt(v)/t  \label{W2v} 
\ea
where $C_i$'s are the small curves around the singularities  in $v$-plane,
and $C_{v,\infty}$ is a large circle around the $v=\infty$. 
Notice that this is the same as the the contribution of the the circle 
near $w=0$  to $W^w_2(\Sigma)$ as it should be, 
since  the circle near $w=0$ is equivalent to the circle near $v=\infty$. 
Therefore by adding $W_2^w$ and $W_2^v$, the contribution  of the circle 
at the asymptotic region of NS brane is weighted by two relative that of 
the tubes. Therefore we we prove that 
\be 
W_3(\Sigma)=W_2^{(v)} +W_2^{(w)} ,\label{relation}
\ee
which gives the relation of 'old' and new super-potential at least for the 
orthogonal NS five branes. This relation 
can be proved case by case in  the examples below.  

Now we give some explicit examples.
The first example is the brane configuration corresponding to the $SU(N)$ 
super Yang-Mill theory.  The curve is given by 
\begin{equation}
        vw=-\zeta ,\;\;  t=v^N 
        \label{pure}
\end{equation}
and we get the value of the super-potential  $   N \zeta$ both in $v$- 
and $w$-pictures. This is 
consistent to the value of SQCD as well as  verifying the relation eq.\ref{relation}.

The second and more interesting example is  the case of $SU(N)$ with $F$
hyper multiplets in the fundamental representation.
The curve is given by 
\begin{equation}
        vw=-\zeta ,\;\;  t=w^{N-F} \prod_{i=1}^F (w-w_i) 
        \label{NandF}
\end{equation}
The value of the super-potential in this case  is 
\be W^{(w)}_2=   \zeta ((N-F)+F)=  \zeta N,
\ee
which agrees with the known value in gauge theory\cite{seiberg-intril}.
Here (N-F) is contribution from 
the infinite circle of NS brane ($w=0$) and $F$ is the  sum of those 
from infinitesimal circles in the tubes ($w=w_i$'s). 
In the v-picture, we get $ (N -F)R\zeta $. 
Notice that in  both  examples, $W_3$, 
the value of the super-potential calculated in generalized Witten's method 
is the sum of the values in electric and magnetic pictures in the new 
definition:  namely,
\be 
W_3(\Sigma)=W_2^{(v)} +W_2^{(w)} = \zeta (2N-F) .\label{relation1}
\ee

One should also notice the  semi-topological character of the super-potential:
The value of 
the super-potential does not depend on the location of the semi-infinite D4 
branes only through the value of  $\zeta$. 
It is consequence of  $wv=-\zeta$, i.e, the orthogonality of the two NS branes. 
This is not surprising, since the value of the super-potential depends only on 
the vacuum configuration where  no  massive excitation are created. 
Therefore we do not expect explicit dependence on the quark masses.  

\section{More examples: The cases including orientifold}
The curves for brane configuration that correspond to  $N=1$ SQCD with 
SO(N), Sp(N) gauge theories were given in \cite{skiba}.
First, Sp(N) with $F$ flavor in fundamental representation, the curve is given by 
\begin{eqnarray}
        v w& = & -\zeta  \no
        t & = & \xi w^{N+2-F}  \prod_{i=1}^{F/2} (w^2-w_i^2),
        \label{spn}
\end{eqnarray}
where 
\be 
\xi=({\rm Pf}M)^{-2(N+2)/F}, ~~{\rm and }~~
\zeta=\Lambda_{Sp}^{(3(N+2)-F)/(N+2-F))} (\det M)^{F/(F-N-2)}. \ee
The value of the super-potential is 
\be
W^w_2=  \zeta  (N+2)
\ee
Notice that this is the asymptotic bending power of the NS brane where the 
 $F$ semi-infinite brane is attached. This is in electric picture.
In the   v-picture,  
\be
W^v_2=  \zeta  (N+2-F)
\ee

Next SO(N) with $F$ flavor in fundamental representation:
The curve is given by 
\begin{eqnarray}
        v w& = & -\zeta  \no
        t & = & \xi w^{N-2-F}  \prod_{i=1}^{F/2} (w^2-w_i^2),
         \label{son}
\end{eqnarray}
where 
\be 
\xi=({\rm Pf}M)^{-2(N-2)/F}, ~~{\rm and }~~
\zeta=\Lambda_{Sp}^{(3(N-2)-F)/(N-2-F))} (\det M)^{F/(F-N+2)}. \ee
 The value of the super-potential  is 
\be
W_2^w=\zeta  (N-2),
\ee
in the meson picture. 
In the v-picture,  
\be
W= \zeta (N-2-F)
\ee
In all of these examples $W_{SQCD}=W_2^w(\Sigma)$ and  the relation 
$W_3=W_2^w+W_2^v$ holds.

Finally we go back to the non-orthogonal case
Now it is time to ask what happen to the non-orthogonal case with the new 
definition of the super-potential.
The curve, $\Sigma$, for the gauge theory with massive quarks 
can be written as 
\begin{eqnarray}
v+m_f &= &{1\over\mu}{(w-w_+)(w-w_-)\over w}                                    \\
t(w) & = & w^{N-F} (w-w_+)^r(w-w_-)^{F-r},       \label{rot}
\end{eqnarray}
Upon the pull back to the w plane, we again have $v(w) dw\wedge dt/t =0$.
\ba
W_2(\Sigma)&&= -{1\over 2\pi i} \int v(w)wdt/t  \no
                &&=    -\frac{w_-w_+}{\mu}(N-F) +m_f{\rm Tr}M  .
\ea
This value is not the same as the value of the gauge theory.
The difference is $ -\frac{w_-w_+}{\mu}N$. We expect that this is the 
contribution from the asymptotic region of the NS five brane, although
it is not clear how to compute the $W^v_2$  in this non-orthogonal case.
It would be very intersting to prove  that this is the case.

\section{discussion}
In this paper we showed that $W_3$, Witten's  definition of the super 
potential,  does not give the same value to that of the gauge theory 
for the orthogonal NS and NS' branes.
We gave a  modified definition 
of MQCD super-potential, $W_2$,  which gives the same value as that of SQCD.
We give the relation between $W_3$ and $W_2$. 
For non-orthogonally rotated brane configuration which corresponds to the 
gauge theory with mass of the adjoint field, $W_3$ fits the value of the 
gauge theory.     
It is still  unclear how to understand the  gauge theory super-potential for the 
non-orthogonal case from the point of new definition. 
Why $W^w_2$ fails for non-orthogonal case? 
It is also interesting to check  the  relation  eq.\ref{relation} continues to hold 
in the presence of the more general tree level super-potential 
$W_{tree}(\Phi)$. These issues are under investigation.

\vskip 1cm
\noindent{\bf \Large Acknowledgement}

\noindent I would like to thank  Ken Intriligator,  
Soonkeon Nam  and Yongsung Yoon for useful discussions. 
This work has been supported by the research grant 
KOSEF 971-0201-001-2.

\end{document}